# AI Agentic workflows and Enterprise APIs: Adapting API architectures for the age of AI agents

Vaibhav Tupe
*Equinix*
Redwood City, USA
vaibhav.k.tupe@gmail.com

Shrinath Thube
*IBM*
San Jose, USA
shrinaththube@gmail.com

*Abstract*—The rapid advancement of Generative AI has catalyzed the emergence of autonomous AI agents, presenting unprecedented challenges for enterprise computing infrastructures. Current enterprise API architectures are predominantly designed for human-driven, predefined interaction patterns, rendering them ill-equipped to support intelligent agents' dynamic, goal-oriented behaviors. This research systematically examines the architectural adaptations for enterprise APIs to support AI agentic workflows effectively. Through a comprehensive analysis of existing API design paradigms, agent interaction models, and emerging technological constraints, the paper develops a strategic framework for API transformation. The study employs a mixed-method approach, combining theoretical modeling, comparative analysis, and exploratory design principles to address critical challenges in standardization, performance, and intelligent interaction. The proposed research contributes a conceptual model for next-generation enterprise APIs that can seamlessly integrate with autonomous AI agent ecosystems, offering significant implications for future enterprise computing architectures.

## I. INTRODUCTION

The proliferation of artificial intelligence (AI) technologies is reshaping enterprise computing, with autonomous AI agents emerging as pivotal entities in modern workflows. These agents, capable of performing complex tasks independently, are transforming how enterprises manage processes, data, and decision-making [1]. However, existing enterprise API architectures are largely tailored for static, human-driven interactions, posing significant challenges in adapting to the dynamic, iterative behaviors of AI agents [2].

Traditional APIs rely on predefined endpoints and structured query responses, optimized for predictable workloads [3]. In contrast, AI agents demand flexibility, context-aware interactions, and real-time adaptability to function efficiently within enterprise ecosystems [4]. This paradigm shift necessitates rethinking API design to accommodate intelligent agents capable of multi-agent collaboration, tool utilization, and continuous learning [5].

This paper aims to explore how enterprise APIs can evolve to support AI agentic workflows. By examining the core characteristics of AI agents, current enterprise API practices, and emerging challenges, we propose a framework for building "agent-ready" APIs. Our approach emphasizes standardization, scalability, adaptability, and security, addressing the unique needs of agent-driven environments and laying the foundation for next-generation enterprise computing architectures [6].

## II. RELATED WORK

### A. Understanding AI Agents and Agentic Workflows

*1) AI Agents: Characteristics and Capabilities:* AI agents are autonomous software entities designed to perceive their environment, process information, and execute actions to achieve specific objectives [4]. These agents operate using advanced machine learning models and algorithms, enabling them to function independently or collaboratively in dynamic environments without constant human supervision [1]. Their ability to adapt and learn from changing contexts makes them a transformative technology across various domains [4].

*a) Core Characteristics of AI Agents:* The core characteristics of AI agents distinguish them from traditional automation systems. Autonomy allows these agents to make decisions and take actions without human intervention, enhancing their capability to handle complex tasks. Adaptability ensures that agents can modify their behavior based on feedback or environmental changes, making them resilient in unpredictable scenarios [4]. Proactivity empowers AI agents to anticipate user needs and initiate actions instead of waiting for instructions, improving efficiency in task execution [5]. Furthermore, their collaborative capability allows agents to integrate seamlessly with other agents and systems, facilitating cooperative problem-solving [7]. These features collectively enable AI agents to surpass traditional automation by providing dynamic responses and continuous self-improvement through iterative processes [6].

*b) Key Capabilities:* AI agents possess key capabilities that make them valuable. Reflection mechanisms enable agents to self-evaluate their actions and refine their performance over time. For instance, models like Reflexion use feedback loops to optimize decision-making processes [4]. Planning capabilities allow agents to structure sequences of actions to achieve their objectives, as demonstrated by systems such as HuggingGPT, which orchestrates multi-step tasks across various tools and services [5, 7]. AI agents also extend their tool utilization by interacting with external APIs, leveraging frameworks like Gorilla to ensure accurate and dynamic API usage [8]. Finally, multi-agent collaboration enables these agents to coordinate tasks across distributed systems. An example is ChatDev, which employs multiple agents working together to manage software development workflows efficiently [9]. These capabilities make AI agents adaptable and capable of addressing complex, real-world challenges in enterprise environments.

*2) Agentic Workflow Architectures:* Agentic workflow architectures define how AI agents interact within a system, utilize resources, and execute tasks while dynamically adapting to environmental changes [10]. These workflows are designed to optimize task execution and decision-making by integrating key elements such as reflection, planning, tool utilization, and multi-agent collaboration [11]. A well-designed agentic workflow ensures that agents can operate autonomously, handle complex tasks efficiently, and improve performance over time [8, 12].

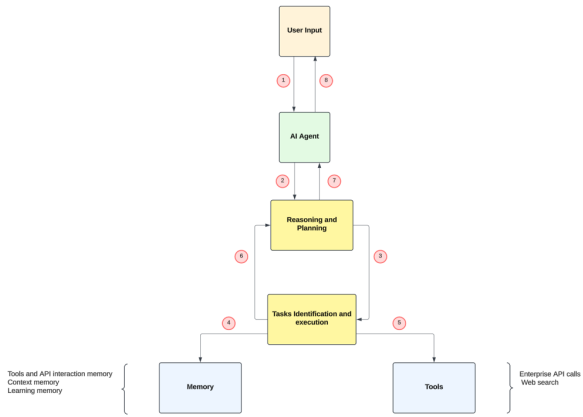

Fig. 1. Overview of the operational flow of an AI agent system [10].

As shown in Fig. 1, the AI agent begins by processing user inputs. The reasoning and planning module identifies the appropriate tasks, which are then executed by leveraging two key components: memory and tools. The memory module stores contextual information, while the tools module handles external interactions, such as API calls and web searches. This feedback loop enables continuous improvement in task execution [10].

B. *Enterprise API's: Current landscape*

Enterprise APIs form the foundation of modern digital ecosystems, enabling seamless communication and integration between diverse systems, applications, and services. They are pivotal in ensuring interoperability, scalability, and innovation within complex enterprise environments [2].

*1) The role of APIs in Enterprise ecosystem:* APIs serve as vital connectors within enterprise ecosystems, facilitating seamless communication and integration between diverse systems and applications. They play a critical role in ensuring interoperability by bridging legacy systems with modern applications, thereby maintaining continuity and ensuring smooth data flow across evolving architectures [2]. APIs also enhance scalability by leveraging standardized frameworks that provide modular, reusable interfaces capable of handling increasing workloads efficiently [3]. Furthermore, APIs are essential in driving innovation, enabling enterprises to quickly deploy new features, integrate third-party services, and build expansive ecosystems around core business applications. Additionally, APIs support integration by utilizing standardized protocols such as REST, GraphQL, and gRPC to connect diverse enterprise systems, such as CRM, ERP, and cloud services, ensuring interoperability across various platforms [3, 13].

*2) Current Trends in Enterprise APIs:* Several emerging trends are reshaping the present and future landscape of enterprise APIs, influencing how businesses develop and manage integrations. The API-first approach has gained traction as enterprises prioritize API development from the outset of software design, ensuring consistent interfaces and enhancing user experiences across platforms [14]. Multi-cloud and hybrid integrations are also becoming increasingly important, as APIs facilitate seamless operations by bridging cloud providers with on-premise systems, enabling distributed environments to function cohesively [10]. Additionally, API monetization is transforming how organizations view APIs, with many enterprises turning them into products and offering them on marketplaces to generate new revenue streams [15]. Finally, the automation of the API lifecycle is streamlining processes, with tools like automated API documentation generators improving consistency, reducing manual effort, and accelerating the overall development lifecycle [11].

*3) Evolution of API design and usage:* The evolution of enterprise APIs marks a significant transition from basic web API designs to more sophisticated and adaptive architectures. RESTful APIs, known for their stateless nature and simplicity, continue to be a foundational aspect of enterprise systems, facilitating scalable and reliable integrations [3]. Beyond REST, event-driven architectures, utilizing Webhooks and streaming protocols, enable real-time interactions, which are vital for applications demanding immediate updates [12]. API gateways play a crucial role in managing these interactions by centralizing control over traffic, ensuring security, and providing monitoring capabilities necessary for large-scale deployments [10]. Additionally, microservices architectures rely heavily on APIs to interconnect modular and scalable components, forming cohesive and adaptable applications [16]. These advancements illustrate a clear progression in API design, aiming to address the growing complexities of modern enterprise ecosystems.

III. METHODOLOGY

This study uses a conceptual research methodology to develop a framework for adapting enterprise APIs to support AI agentic workflows. It involves two steps: theoretical analysis and model development. The first step reviews literature on AI agents and API architectures, identifying gaps in traditional API designs through comparative analysis. Based on these insights, the study proposes an agent-ready framework that includes API standardization, context awareness, adaptive responses, and developer support to meet the evolving needs of intelligent agents.

IV. CHALLENGES FOR ENTERPRISE APIs IN AGENTIC WORKFLOWS

Agentic workflows demand advanced API capabilities to enable intelligent agents to perform dynamic, iterative, and collaborative tasks. However, adapting traditional APIs to meet these requirements introduces significant challenges.

A. *Scalability and High-Performance demands*

Agentic workflows, characterized by iterative refinement processes and multi-agent collaborations, significantly increase API traffic, posing several performance challenges. One major issue is dynamic workloads, where APIs must handle fluctuating demands as agents continuously iterate over large datasets [3]. For instance, Gorilla LLM highlights how even minor API misuse, such as unnecessary large payload requests, can cause performance bottlenecks [8]. Another challenge is latency sensitivity, particularly for real-time and critical applications that require sub-second API response times. Maintaining such low latency under high loads is difficult, yet essential for ensuring seamless interactions in time-sensitive use cases [8].

*B. Inadequate API Flexibility*

Static API designs often fail to meet the dynamic demands of agentic workflows. One major limitation is rigid payloads, where intelligent agents require real-time adjustments to API responses based on context, but most APIs are not built to handle mid-query modifications [4]. Another challenge is versioning overhead, where maintaining multiple API versions to support different agent behaviors increases technical debt and reduces flexibility [12]. To enable efficient agent interactions, APIs need to evolve into more adaptable, context-aware systems that can adjust to changing requirements without relying on rigid structures or excessive versioning [5, 10].

*C. Security and Compliance Risks:*

As APIs increasingly interact with intelligent agents handling sensitive data, ensuring robust security measures is essential [12]. Dynamic authentication is a key challenge, as traditional token-based or role-based systems often fail to adapt when agents dynamically request access to sensitive operations [10]. Additionally, data protection is critical, requiring APIs to comply with regulations like GDPR and HIPAA while still providing agents with contextually relevant information [11]. Balancing security, privacy, and usability is necessary to protect sensitive data without hindering agent-driven workflows [12].

*D. Real-Time Collaboration among Agents*

Agentic workflows often involve collaborative agents that exchange information and coordinate tasks. However, poorly designed APIs can introduce synchronization delays, especially in multi-cloud environments, when agents struggle to align data across systems. Additionally, without proper orchestration, agents may make redundant requests, leading to inefficient use of bandwidth and computing resources [7, 9]. These issues can hinder the performance of agentic workflows, making it crucial to optimize API designs for better synchronization and resource management.

*E. Poor API documentation and Usability*

Poor API documentation can hinder the performance of intelligent agents. Ambiguity in endpoints, as seen in Gorilla's framework, leads to "hallucination errors" where agents misinterpret unclear documentation, resulting in invalid API calls [8]. Additionally, insufficient metadata prevents agents from making informed decisions due to a lack of critical details like data ranges and query limits [8]. Clear documentation and comprehensive metadata are essential to support accurate agent interactions.

*F. Inefficient Query Optimization*

Agents often generate broad or redundant queries, consuming excessive resources and reducing performance, as seen in Gorilla's agent-API evaluation [8]. This leads to cost implications in pay-per-call APIs, where unnecessary requests inflate infrastructure and computing costs.

*G. Challenges with Legacy Systems*

Legacy systems face protocol compatibility issues with modern agentic workflows due to outdated protocols [3]. Additionally, resource constraints in these systems, such as limited computational and network capabilities, can hinder agents' performance by slowing operations and reducing efficiency.

*H. Governance and Monitoring*

Effective governance and monitoring are crucial for managing agentic workflows. However, many APIs suffer from a lack of transparency, as they are designed for static use cases and lack the necessary monitoring and logging mechanisms to track agent interactions. Additionally, policy enforcement is complex in decentralized systems, making it difficult to apply granular API usage policies across multiple agents and workflows. Strengthening governance frameworks is essential to ensure secure and efficient agent operations [12].

## V. STRATEGIC APPROACH OF API ADOPTION

Adapting enterprise APIs to the demands of agentic workflows requires a strategic shift in design, implementation, and management. This section outlines actionable strategies to enable APIs to support intelligent agents effectively, focusing on scalability, adaptability, and collaboration.

*A. Agent Specific API Standardization*

While existing standards like REST, GraphQL, and gRPC facilitate human-driven and predefined application workflows, they fall short in addressing the unique requirements of autonomous agents. This gap necessitates the establishment of agent-specific API standards that cater to the distinct capabilities and operational needs of AI agents.

*1) Intent Based API Endpoint design:* To optimize API interactions for AI agents, intent-based endpoint design allows agents to perform complex tasks through high-level actions rather than multiple granular calls. Instead of traditional CRUD endpoints, intent-based APIs interpret the agent's intent. This enables the backend to process broader operations in a single call, improving efficiency and reducing unnecessary request cycles. For example, an endpoint like "/order/manage" can dynamically handle various tasks such as creating, updating, or canceling an order based on the intent received. This design supports adaptive workflows by simplifying interactions and enabling AI agents to achieve task-specific goals through minimal requests.

*2) Agent Specific API Header:*

*a) Context IDs for Session Continuity:* Enables agents to track ongoing interactions across multiple API calls, reducing redundant queries.

*b) Intent-Based Headers:* Embeds agent intent directly in API headers (e.g., "X-Agent-Intent: OrderStatusCheck"), minimizing endpoint calls.

*c) Agent Role Identifiers*: Standardizes headers to tailor API responses to specific agent roles like customer service or analytics.

*d) Human vs. AI* Interaction Differentiation: Uses headers (e.g., "X-Agent-Type: AI") to distinguish AI agents from human users, enabling custom API policies.

*e) Token Claims for Agent Verification*: Requires tokens that confirm requests originate from AI agents, enhancing security and access control.

*f) Timestamp Metadata for Data Freshness:* Provides headers like "X-Data-LastUpdated" to indicate when the data was last updated, helping agents make decisions based on data relevance.

*g) Rate-Limit and Error Recovery Headers*: Includes headers like "X-RateLimit-Remaining" to manage usage

quotas and "X-Error-Recovery: RetryAfter=60s" to guide agents on efficient error handling and retries.

These API Header enhancements improve interoperability and enable efficient, context-aware responses that streamline agent workflows. Differentiating AI agents from human users allows APIs to enforce custom rate limits, intent verification, and secure data access, preventing misuse and ensuring scalable, reliable interactions in agent-driven ecosystems.

*3) Metadata Improvement*: Embedding dynamic metadata in API responses enables AI agents to adapt workflows in real time. Rich metadata, such as data freshness indicators and error recovery suggestions, reduces redundant queries, optimizes decision-making, and supports efficient multi-turn interactions, ensuring more context-aware and adaptive API usage.

*4) Agent Query Language:* An Agent Query Language (AQL) is essential for standardizing how AI agents interact with enterprise APIs, enabling more efficient and context-driven data retrieval. Unlike traditional request models, AQL should support intent-based querying, allowing agents to ask high-level, goal-oriented questions that APIs can interpret dynamically. Additionally, GraphQL-style queries can be adopted to allow agents to request only the necessary fields, minimizing payload size and reducing redundant data transfer. A standardized AQL empowers agents to make adaptive, context-aware requests, improving API interactions and reducing backend load by consolidating multiple queries into a single, intelligent request.

*5) API Documentation:* Traditional API documentation often caters to human developers, but AI agent-aware API documentation must be machine-readable, dynamic, and discoverable to optimize agent workflows. Documentation formats such as OpenAPI and GraphQL introspection should be utilized to enable agents to programmatically discover available endpoints, query structures, and required parameters. Furthermore, API documentation should include intent-based usage guides and metadata descriptions to help agents understand how to use context-specific headers and response formats effectively [13, 16]. For example, an agent could query an "/api/discover" endpoint to retrieve updated documentation in real-time, ensuring the API's evolving capabilities are continuously available. This dynamic and interactive documentation approach allows agents to self-adapt to new API functionalities, reducing manual intervention and improving automation in multi-turn interactions.

*B. Stateful Context Aware Middleware*

To enable AI agents to deliver context-aware responses, enterprises should integrate middleware that manages state and context across API interactions. This context-aware middleware acts as an intermediary between stateless API endpoints and the AI agent, retrieving relevant user session data and appending it to incoming requests. By decoupling context management from the API itself, the middleware ensures that API endpoints remain scalable and maintainable while supporting stateful interactions. This approach reduces redundant queries and improves agent performance by dynamically restoring interaction history, thereby enhancing response accuracy and continuity.

*C. Scalability and Performance*

Implementing auto-scaling infrastructure and load balancing ensures APIs can dynamically adjust to traffic spikes without performance degradation. To further enhance performance, two key aspects need to be considered:

*1) Queue Management for Multi-Turn Interactions:* AI agents often engage in multi-step workflows, requiring APIs to manage these interactions efficiently. Introducing priority-based queues ensures time-sensitive requests are processed faster, while intelligent queue management prevents resource overloading by handling interactions asynchronously. This approach is essential for ensuring smooth, uninterrupted conversations in agent-driven workflows.

*2) Context-Aware Caching:* AI agents frequently request personalized or repetitive data, making caching crucial for reducing backend load. Implementing context-aware caching allows responses to be stored and reused based on session context and intent, minimizing redundant queries. For instance, caching order summaries during a session reduces repeated data retrieval and improves response time for subsequent requests.

Additionally, payload optimization techniques such as AQL queries allow agents to request only the necessary fields, minimizing response size and improving network efficiency. Asynchronous request handling and retry policies further ensure APIs remain scalable and resilient, even under unpredictable agent-driven workloads.

*D. Monitoring, Security and Compliance*

Securing APIs for AI agents requires implementing agent-specific security policies to control data access, ensure compliance with privacy regulations, and prevent misuse by autonomous agents. Below are the key considerations for securing AI agent interactions:

*1) Defining Agent Roles and Access Policies:* AI agents operate autonomously and perform diverse tasks such as data retrieval, user interactions, or transactional operations. APIs should enforce role-based access control (RBAC) to define agent-specific roles (e.g., Support Agent, Analytics Agent, Order Processing Agent) and grant access based on these roles. Fine-grained authorization policies using OAuth 2.0 scopes (e.g., order: read, profile: update) should be applied to ensure agents only perform tasks aligned with their designated roles.

*2) Differentiating Human and Agent Interactions:* This distinction allows APIs to enforce custom rate limits, intent verification, and access restrictions for agents, preventing misuse and optimizing performance.

*3) Dynamic Consent Management:* AI agents often handle sensitive user data, requiring APIs to comply with privacy regulations such as GDPR and CCPA. Providing dynamic consent management endpoints (e.g., /consent/update) allows agents to request and update user permissions in real-time, ensuring compliance with user data preferences.

*4) Audit Logging and Anomaly Detection:* To maintain compliance and detect potential misuse, APIs must log agent interactions including intent, request context, and data accessed. Implementing real-time anomaly detection can help monitor unexpected agent behaviors, such as unauthorized data access attempts or excessive retries, reducing the risk of abuse.

By defining agent-specific roles and enforcing agent-aware policies, enterprises can secure their APIs while ensuring agents interact responsibly within regulatory boundaries and system constraints.

*E. Agent Development Kit (ADK)*

To enable developers to create robust AI agents that seamlessly interact with agent-aware APIs, an Agent Development Kit (ADK) should provide a set of tools, templates, and best practices. Below are key considerations for providing a comprehensive ADK:

*1) Prompt Playbook and Replay:* Provide a Prompt Playbook containing predefined prompt templates and common agent queries to help developers build effective conversations. The Replay feature should enable agents to replay prior interactions for testing, debugging, and optimization, ensuring agents handle multi-turn interactions accurately.

*2) Agent Testing Sandbox:* Include an Agent Testing Sandbox to allow developers to test agent interactions in a controlled environment. The sandbox should simulate various API responses, edge cases, and error scenarios, enabling agents to be tested against real-world conditions before deployment.

*3) Intent Templates and Query Builders:* Provide ready-made intent templates and query builders that align with the intent-based API design. These tools simplify agent queries and ensure agents interact with APIs in a structured and compliant manner.

*4) Monitoring and Governance Libraries:* Include monitoring tools that track agent performance metrics such as intent success rates and error frequency. The ADK should also support real-time anomaly detection to ensure agents remain compliant with governance policies.

By offering a comprehensive ADK, organizations can empower developers to build intelligent, efficient, and compliant AI agents that are optimized to interact with enterprise-grade APIs. This approach simplifies agent development, enhances agent reliability, and improves overall API utilization.

*F. Proposed Architectural Framework*

The architecture in Fig. 2 addresses key AI agent-specific considerations to optimize API interactions. An Edge Cache/CDN reduces latency by caching frequently accessed responses, ensuring faster request handling for AI agents. The API Gateway enforces agent-specific roles, permission policies, and zero-trust security, which are critical to control and secure AI-driven requests. Additionally, it provides custom rate limiting and usage monitoring tailored to the unique behavior of AI agents, preventing misuse and managing API consumption. GraphQL Federation enhances data retrieval by allowing agents to request precise datasets from multiple microservices via a unified endpoint, avoiding over-fetching or under-fetching of information [13, 16]. The architecture also supports dynamic data relationship handling, enabling the agent to efficiently query complex data structures. These API-level enhancements ensure that the system can handle the dynamic and high-frequency request patterns of AI agents, while maintaining scalability, security, and performance.

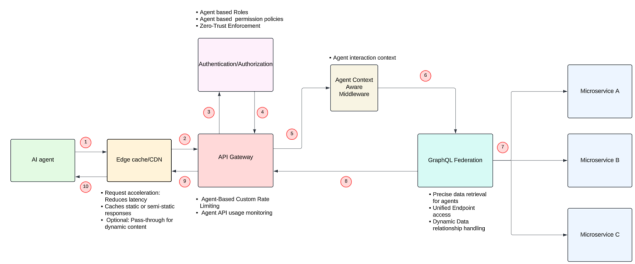

Fig. 2. AI agent-aware enterprise API design architecture considerations.

## VI. CHALLENGES AND FUTURE WORK

Despite the advancements proposed in API adoption design, several challenges remain unresolved, which future research must address to ensure seamless integration of AI agents within enterprise APIs. One significant challenge is the standardization of agent-API communication. Currently, the lack of universal standards for how agents interact with APIs hinders interoperability across diverse systems and platforms. The traditional OpenAPI/Swagger specifications, while effective for human-driven interactions, fall short in accommodating the dynamic and context-aware nature of AI agents. There is an urgent need to evolve these standards to incorporate AI-friendly descriptions of business logic, constraints, and expected behaviors, thereby enabling agents to better interpret and utilize APIs effectively.

Another pressing concern is agent behavior governance. As AI agents become more autonomous, robust monitoring and governance frameworks are essential to manage and audit their interactions with enterprise APIs. Enterprises must be able to track agent activities, enforce compliance policies, and ensure that agents adhere to expected behaviors in dynamic environments. This includes incorporating real-time observability tools and audit trails to enhance transparency, detect anomalies, and mitigate potential misuse of API functionalities.

Additionally, balancing stateless design principles with the contextual awareness needs of AI agents presents a novel architectural challenge. While stateless APIs are preferred for scalability and simplicity, AI agents often require contextual information to make more informed decisions [5, 7]. Future API designs must find ways to reconcile this conflict by supporting context-sharing mechanisms without compromising the benefits of statelessness. Furthermore, declaring dependencies between operations is another area requiring attention, especially when agents need to orchestrate complex workflows involving multiple interdependent API calls.

Lastly, new testing approaches for AI-specific interaction patterns are critical. Traditional testing frameworks focus on deterministic scenarios, but AI agents often exhibit non-deterministic behavior based on dynamic inputs and environmental factors [5, 8]. Future work must develop testing methodologies that account for these variables, ensuring that agent-API interactions remain robust, secure, and efficient under various conditions. These advancements will collectively contribute to creating a resilient ecosystem where enterprise APIs can effectively support the next generation of AI-driven workflows.

## VII. Conclusion

The rise of AI-driven agentic workflows represents a paradigm shift in enterprise operations, necessitating a fundamental evolution in API design and management. This paper synthesizes key findings, demonstrating how agent-ready APIs can enable seamless integration, adaptability, and scalability to support intelligent agents across diverse domains.

Agent-ready APIs are transformative, providing the foundation for dynamic, context-aware interactions, real-time multi-agent collaboration, and robust security frameworks. By adopting strategies such as context-aware designs, event-driven architectures, and AI-augmented functionalities, enterprises can unlock unprecedented levels of efficiency and innovation. These adaptations not only optimize existing workflows but also pave the way for entirely new capabilities in decision-making, task orchestration, and cross-domain interoperability.

As intelligent agents increasingly drive enterprise workflows, the role of APIs becomes more critical than ever. This paper calls on the industry to adopt the strategies and guidelines discussed, fostering a culture of innovation that embraces the potential of agent-ready APIs. Simultaneously, the research community must continue exploring adaptive API designs, explainability mechanisms, and sustainable practices to address emerging challenges and opportunities.

The future of intelligent, agent-driven ecosystems depends on the readiness of APIs to meet the demands of these workflows. By taking proactive steps today, enterprises and researchers can shape a resilient, scalable, and ethical digital infrastructure for the AI-driven era.